\documentclass[12pt]{iopart}
\usepackage{graphicx}
\usepackage{multirow}
\usepackage{hyperref}
\usepackage{iopams}
\usepackage{cite}
\usepackage{color}
\newcommand{\figurewidth}{0.6\textwidth}

\begin{document}

\title[Stepwise contraction of the nf shells in the 3d photoionization of Xe$^{3+,4+,5+}$ ions]{Stepwise contraction of the nf Rydberg shells in the 3d photoionization of multiply-charged xenon ions}

 \author{S.~Schippers$^1$, A.~Borovik Jr.$^1$, T.~Buhr$^2$, J.~Hellhund$^1$, K.~Holste$^1$, A.~L.~D.~Kilcoyne$^3$, S.~Klumpp$^4$, M.~Martins$^4$, A.~M\"{u}ller$^1$, S.~Ricz$^5$, S.~Fritzsche$^{6,7}$}

 \address{$^1$ Institut f\"{u}r Atom- und Molek\"{u}lphysik, Justus-Liebig-Universit\"{a}t Gie{\ss}en, 35392 Gie{\ss}en, Germany}
 \address{$^2$ Physikalisch-Technische Bundesanstalt, 38116 Braunschweig, Germany}
 \address{$^3$ Advanced Light Source, MS 7-100, Lawrence Berkeley National Laboratory, Berkeley, California 94720, USA}
 \address{$^4$ Institut f\"{u}r Experimentalphysik, Universit\"{a}t Hamburg, 22761 Hamburg, Germany}
 \address{$^5$ Institute of Nuclear Research of the Hungarian Academy of Sciences, Debrecen, P.O. Box 51, H-4001, Hungary}
 \address{$^6$ Helmholtz-Institut Jena, 07743 Jena, Germany}
 \address{$^7$ Theoretisch-Physikalisches Institut, Friedrich-Schiller-Universit\"{a}t Jena, 07743 Jena, Germany}

\ead{stefan.schippers@physik.uni-giessen.de}

\begin{abstract}
Triple photoionization of Xe$^{3+}$, Xe$^{4+}$ and Xe$^{5+}$ ions has been studied in the energy range 670--750~eV, including the $3d$ ionization threshold. The photon-ion merged-beam technique was used at a synchrotron light source to measure the absolute photoionization cross sections. These cross sections exhibit a progressively larger number of sharp resonances as the ion charge state is increased. This clearly visualizes the re-ordering of the $\epsilon f$ continuum into a regular series of (bound) Rydberg orbitals as the ionic core
becomes more attractive. The energies and strengths of the resonances are extracted from the experimental data and are further analyzed by relativistic atomic-structure calculations.
\end{abstract}

\submitto{\JPB}
\maketitle

\section{Introduction}

One of the great successes of atomic physics is its fundamental understanding of the chemical properties of the elements and the explanation of their order within the periodic table. Following the atomic number of elements, the (atomic) subshells are usually filled for each principal quantum number $n$ from the inner to the outer subshells. Prominent exceptions are the $3d$ and $4d$ transition elements, and even more so the lanthanide and actinide groups of elements where the $4f$ and $5f$ subshells are filled only after shells of the next principal quantum number(s) have been (partially) occupied already. For the lanthanides, especially, this irregularity can be explained by the peculiar shape of the potential of $f$-electrons which consists of an inner and an outer well, and which are separated by a centrifugal barrier. For low nuclear charges, the inner well is too shallow to confine the $f$ electrons, so that they are mainly localized in the outer well. For this reason, the $f$-shell behaves first like a diffuse outer shell and remains unfilled until the nuclear charge, and hence the potential, is sufficiently attractive to strongly bind the $f$-electrons.

This contraction of the electron density has been addressed in the literature also as the \lq\lq collapse of the $4f$ wave function\rq\rq\ and has been explored since the early days of atomic-structure theory \cite{GoeppertMayer1941,Griffin1969}. For xenon and its neighbour elements, in particular, the giant resonance in the photoionization cross section \cite{Lucatorto1981,OSullivan1996,Koizumi1997,Bizau2001,Cummings2001,Kjeldsen2002b,Habibi2009} was shown to be due to a $4d-4f$ excitation and the quite sudden change in the spatial overlap of the wave functions, if a $4d$ electron is excited \cite{Karaziya1981,Connerade1982a,Nuroh1982,Cheng1983,Hansen1989}.

An analogue contraction for a whole series of $nf$ orbitals is demonstrated in this work for the $3d$ photoionization of Xe$^{q+}$ ions with charge states $q = 3,4,5$. Until the present, this collapse of the $nf$ series has been investigated only in very few studies on the photoabsorption by or ionization of the inner $3d$ shell, mainly because of a lack of photon sources with sufficient intensity in the required range of energies. For the same reason only neutral targets were considered so far \cite{Sonntag1984,Arp1999,Kivimaeki2000,Richter2007,Fritzsche2008}, which can be prepared with much higher particle densities than ionic targets. A notable exception ist the work of Ma\u{\i}ste et~al.~\cite{Maiste1980} who studied low-charged ions embedded in the lattices of ionic crystals.

\section{Experiment}\label{sec:exp}

The present experimental measurements, with multiply charged ions, employed the photon-ion merged-beam method using the \underline{P}hoton-\underline{I}on spectrometer at \underline{PE}TRA\,III (PIPE) \cite{Schippers2014}. This is a permanently installed end station at the "Variable Polarization XUV Beamline" (P04) \cite{Viefhaus2013} of the world's brightest 3$\mathrm{rd}$ generation synchrotron light source at present, PETRA\,III at DESY in Hamburg, Germany. A  description of the experimental setup and the data-analysis procedures has been reported \cite{Schippers2014}. In the present experiment, xenon ions in charge states $q=3,4,5$ were produced in an electron-cyclotron resonance (ECR) ion source and accelerated to energies of $q\times6$~keV. After the mass/charge selection in a magnetic dipole field, the ions were electrostatically guided onto the photon beam axis. The current of the well collimated ion beams in the merged-beam interaction region ($\sim 1.7$~m length) was 40~nA, 11~nA and 6~nA for the $^{132}$Xe$^{3+}$, $^{132}$Xe$^{4+}$ and $^{132}$Xe$^{5+}$ ions, respectively. The photon flux as measured with a calibrated photo diode amounted up to $3\times10^{13}$~s$^{-1}$ in the 670--770~eV photon energy range of the present experiment. A second magnet, after the interaction region, separated the product ions from the primary beam. The primary ions were directed into a Faraday cup, and the product ions were counted with nearly 100\% efficiency by a single-particle detector.

The photon energy scale was calibrated with an uncertainty of better than $\pm30$~meV by remeasuring known resonances in C$^{3+}$ \cite{Mueller2009a,Mueller2014}. The experimental photon energy spread was 160~meV, i.e., a factor of $\sim$ 20 lower than in our previous study \cite{Schippers2014}. The systematic uncertainty of the cross section scale is $\pm15\%$ \cite{Schippers2014}.

\section{Results and Discussion}\label{sec:res}

\begin{figure}
\centering{\includegraphics[width=\figurewidth]{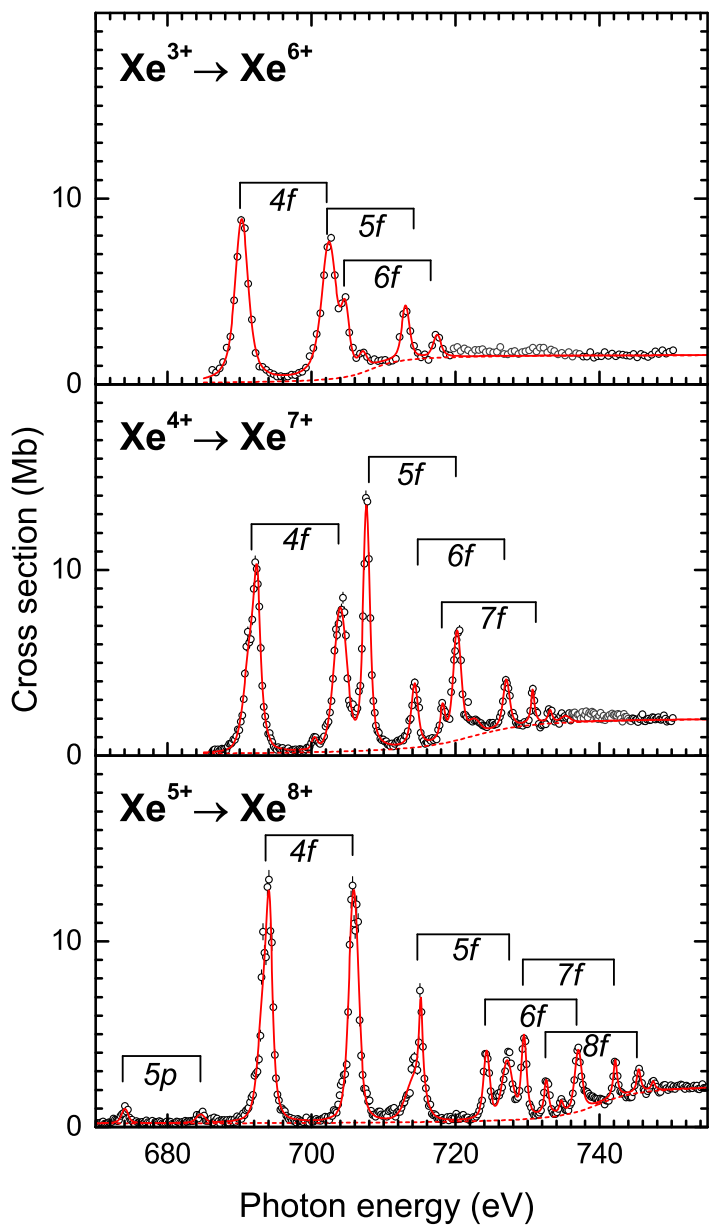}}
\caption{\label{fig:hires} Measured absolute cross sections (symbols) for triple ionization of Xe$^{3+}$, Xe$^{4+}$, and Xe$^{5+}$ ions. The experimental photon energy spread was $\Delta E = 0.16$~eV. The full lines are results of Voigt line profile fits to the measured spectra. The dashed lines are the fitted direct ionization (DI) contributions to the measured cross sections. Resonances are labelled by the $n\ell$ subshell to where the $3d$ electron is excited. Because of the fine structure of the $3d$ hole, there are two resonance features for each $nl$, associated with $j=5/2$ (lower resonance energies) and $j=3/2$ (higher resonance energies), respectively.}
\end{figure}

In our previous low-resolution study on $3d$ photoionization of multiply charged xenon ions \cite{Schippers2014} we found that product ions were created in a range of charge states. The relative contribution of each product charge state depends on the cascade of Auger and radiative processes that follow the initial inner-shell ionization event.   Net triple ionization, i.e., emission of two electrons by autoionization in addition to the directly ionized electron, was found to be one of the strongest channels. Therefore, in the present high-resolution measurements, where the emphasis is on the spectroscopic aspect of the experiment, only the triple ionization channel was investigated.

Measured cross sections for triple photoionization of Xe$^{3+}$, Xe$^{4+}$, and Xe$^{5+}$ are displayed in Fig.~\ref{fig:hires}. Pronounced resonance structures are observed for all ions but with an increasing number of resonances as the charge state of the ions is increased. In contrast to the usually rather complex cross sections for outer shell ionization of atoms and ions, the present inner-shell ionization cross sections can be interpreted straight-forwardly. The strongest resonances are associated with the photoexcitation of a $3d$ electron to an atomic $nf$ subshell ($n \,=\, 4,5,6, ...$) and the subsequent multiple autoionization of the associated hole states. The $3d^{-1}_{3/2}-3d^{-1}_{5/2}$ fine structure splitting ($\sim13$~eV) of the $3d$ hole leads to two distinct Rydberg series of resonances in each spectrum. For Xe$^{3+}$, for example, resonances with principal quantum numbers $n$ from 4 to 7 can be clearly discerned. For Xe$^{5+}$ the series of $nf$ resonances could be observed up to $n=9$, while weaker resonances associated with $3d\to np$ excitations are barely measurable. Only the $3d\to5p$ resonances in the Xe$^{5+}$ spectrum at energies below 685~eV could be measured. For Xe$^{3+}$ and Xe$^{4+}$ no data were taken at these energies.

As seen in Fig.~\ref{fig:hires}, the resonance positions shift towards higher energies as the charge state of the xenon ions is increased. This is attributed to the fact that the binding of the electron is enhanced when successive outer electrons are removed from the xenon atom (reduced screening). The measured resonance widths are significantly larger than the 160~meV experimental energy spread. These are mainly determined by the short lifetime of the $3d$ hole which is rapidly filled by autoionizing transitions from higher atomic subshells. For neutral xenon the associated resonance width is $\sim 0.6$~eV \cite{Arp1999}, and this is consistent with the present observations, although a slight (experimentally unresolved) dependence on the charge state of the ions can be expected. In addition to the natural line width, there is also a resonance broadening due to the fine structure of the $3d^{-1}_{5/2}\,5p^{6-q}\,n\ell$  and  $3d^{-1}_{3/2}\,5p^{6-q}\,n\ell$ hole-state configurations. Some of the measured $4f$ and $5f$ resonance line shapes indicate an associated substructure. For higher $n$ the fine-structure splitting becomes smaller than the natural line width such that individual resonance levels cannot be resolved.

\begin{table}
\caption{\label{tab:res} Resonance energies (in eV) as extracted from the high-resolution spectra in Fig.~\ref{fig:hires}. In addition, we list the quantum defects $\delta$ and Rydberg series limits $E_\infty$ from the fit of the Rydberg formula (Eq.~\ref{eq:ryd}) to the tabulated resonance positions. The values for $E_\infty^\mathrm{(DF)}$ are the results of the present MCDF calculations.}
  \begin{indented}
    \item[] \begin{tabular}{@{}llll}
    \br
     assignment & Xe$^{3+}$ & Xe$^{4+}$ & Xe$^{5+}$ \\
    \mr
 $3d_{5/2}^{-1}5p$        &           &           & 674.1(1)  \\
 $3d_{3/2}^{-1}5p$        &           &           &  684.6(2)   \\
                           &           & 700.40(8)$^\dag$ & \\
 \br
 $3d_{5/2}^{-1}4f_{5/2}$  & \multirow{2}{*}{690.28(3)}
                                      & 691.73(7)  & 693.5(1) \\
 $3d_{5/2}^{-1}4f_{7/2}$  &           & 692.43(4)  & 694.13(7) \\
 $3d_{5/2}^{-1}5f_{5/2}$  & \multirow{2}{*}{702.40(5)*}
                                      & \multirow{2}{*}{707.62(1)}  & 714.3(2) \\
 $3d_{5/2}^{-1}5f_{7/2}$  &           &            & 715.17(4) \\
 $3d_{5/2}^{-1}6f$        & 704.68(7) & 714.31(3)  & 724.31(2) \\
 $3d_{5/2}^{-1}7f$        & 707.1(1)  & 718.20(4)  & 729.50(3) \\
 $3d_{5/2}^{-1}8f$        &           & 720.21(2)* & 732.61(3) \\
 $3d_{5/2}^{-1}9f$        &           & 722.6(2)   &  734.70(7) \\
\mr
 $\delta$           & \phantom{00}1.0(1) &  \phantom{00}0.83(3) & \phantom{00}0.78(3)  \\
 $E_\infty$              & 714(2)   & 727.1(3)   & 742.1(1)  \\
 $E_\infty^\mathrm{(DF)}$& 712.5      & 726.5      & 741.5  \\
\br
 $3d_{3/2}^{-1}4f$       & 702.40(5)*  & 703.99(2)  & 705.86(3) \\
 $3d_{3/2}^{-1}5f$       & 713.05(5)  & 720.21(2)* & 727.17(4) \\
 $3d_{3/2}^{-1}6f$       & 717.4(1)   & 727.05(4)  & 737.04(3) \\
 $3d_{3/2}^{-1}7f$       &            & 730.73(5)  & 742.18(3) \\
 $3d_{3/2}^{-1}8f$       &            & 733.1(1)   & 745.45(5) \\
 $3d_{3/2}^{-1}9f$       &            &  735.7(3)  & 747.5(1)\\
 \mr
 $\delta$   & \phantom{00}0.99(3) & \phantom{00}0.96(1) & \phantom{00}0.85(1)  \\
 $E_\infty$             & 726.5(4)    & 740.7(2)   & 755.3(2)  \\
 $E_\infty^\mathrm{(DF)}$& 725.6      & 739.5      & 754.6  \\
 \br
 \end{tabular}
 \item[]$^\dag$ not identified
 \item[]* unresolved line blend
 \end{indented}
\end{table}

In order to extract the positions and strengths of the resonances, we have fitted line profiles to the measured cross sections. Individual (resonance) features in the spectra were represented by up to two Voigt line profiles with the Lorentzian widths fixed to 0.6 eV. In addition, a continuous cross section due to $3d$ direct ionization (DI) was modelled by an inverse tangent function. This simple approach neglects the fine-structure splitting of the threshold, Nevertheless, it should be sufficiently accurate for the purpose of extracting resonance parameters from the experimental PI cross sections. The fitted DI contribution and the sum of DI and all resonances are shown as dashed and full lines in Fig.~\ref{fig:hires}, respectively. The resonance energies obtained from the fit are listed in Tab.~\ref{tab:res}.

\begin{figure}
\centering{\includegraphics[width=\figurewidth]{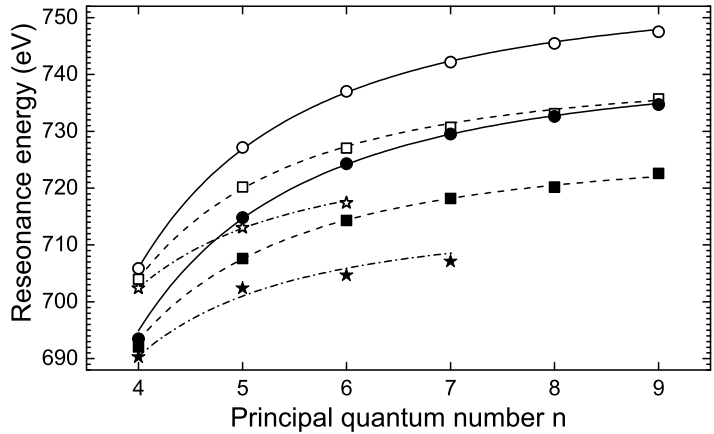}}
\caption{\label{fig:serfit} Resonance energies as obtained from a fit of the Rydberg formula (Eq.~\ref{eq:ryd}) to the measured resonance positions, listed in Tab.~\ref{tab:res}. Full and open symbols refer to the $3d^{-1}_{5/2}\,5p^{6-q}\,n\ell$  and, respectively,  $3d^{-1}_{3/2}\,5p^{6-q}\,n\ell$ resonances of Xe$^{3+}$ ($q=3$, stars), Xe$^{4+}$($q=4$ squares), and Xe$^{5+}$ ($q=5$, circles).}
\end{figure}

As mentioned above, two Rydberg series of $nf$ resonances can be clearly distinguished. They are associated with the two fine structure components of the $3d$ hole. For each of these series, the resonance positions will approximately follow the simple Rydberg formula
\begin{equation}\label{eq:ryd}
    E_\mathrm{res} = E_\infty - \mathcal{R}\frac{(q+1)^2}{(n-\delta)^2}
\end{equation}
with $q$ denoting the primary ion charge state. We have fitted this formula (with $\mathcal{R}\approx 13.606$~eV) to the experimental $nf$ resonance positions (Fig.~\ref{fig:serfit}) in order to extract the series limits $E_\infty$ and quantum defects $\delta$ as displayed in Tab.~\ref{tab:res}. The quantum defects range from 0.8 to 1.0 and show that the $3d_{3/2,\,5/2}$ holes increasingly overlap with the (radial extent of the) $4f$ wave function as the charge state of the xenon ion increases. The series limits  are in reasonable agreement with results from our atomic structure calculations (see below). The series limits also coincide with the thresholds for $3d$ DI. Indeed, the onsets of the DI cross sections in Fig.~\ref{fig:hires} (dashed lines) agree with the $3d^{-1}_{5/2}\,nf$ series limits as well as one might expect from the crude fit model for the DI cross section. For each charge state, the difference between the $3d^{-1}_{3/2}\,nf$ and $3d^{-1}_{5/2}\,nf$ series limits corresponds to the fine-structure splitting of the $3d$ hole. Our analysis gives rise to the values of 12(2), 13.6(4), and 13.2(2)~eV for Xe$^{3+}$, Xe$^{4+}$, and Xe$^{5+}$, respectively. These values agree with each other within the experimental uncertainties and are also close to the value of 12.7~eV for neutral xenon \cite{Larkins1977}, indicating that the number of outer shell $5p$  electrons does not have a strong influence on the structure of the $3d$ shell.

\begin{figure}
\centering{\includegraphics[width=\figurewidth]{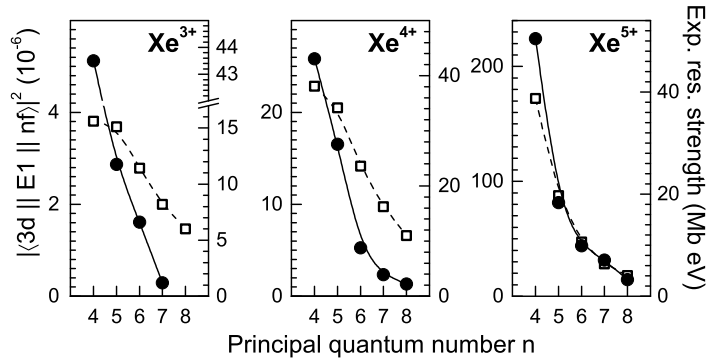}}
\caption{\label{fig:strength} Comparison of the squared moduli of the $3d\to nf$ electric-dipole matrix elements (atomic units, open symbols, left scales) with the experimental resonance strengths (full symbols, right scales). For each $n$, the sum of the experimental  strengths of the corresponding $3d_{5/2}^{-1}\,nf$ and $3d_{3/2}^{-1}\,nf$ resonances is plotted. The lines are just drawn to guide the eye.}
\end{figure}

\begin{figure}[t]
\centering{\includegraphics[width=\figurewidth]{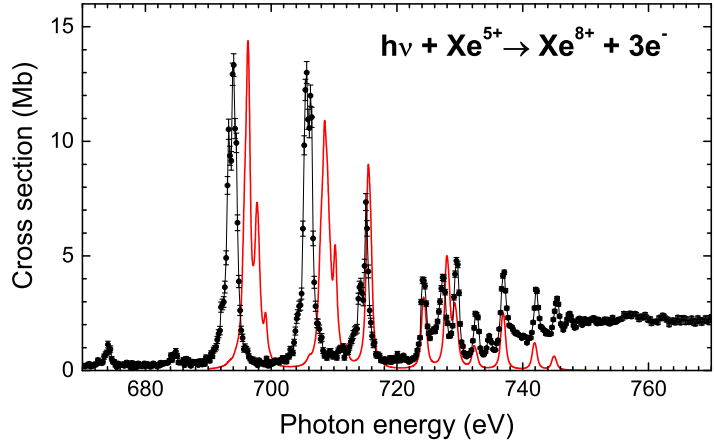}}
\caption{\label{fig:xe5} Experimental (symbols) and theoretical (full line) cross section for the triple photoionization of Xe$^{5+}$.}
\end{figure}

In contrast to the $3d$ fine structure splitting, the formation of the $nf$ Rydberg shells is found to be much more strongly affected by the ion charge state. Additional higher $nf$ orbital functions seem to collapse into (or, at least, towards) the ionic core, i.e., into the inner potential well. The resulting enlarged overlap of the $nf$ with the $3d$ wave functions also leads to enhanced probabilities for resonant $3d\to nf$ excitations, as it is readily seen in the spectra of Fig.~\ref{fig:hires}. Indeed, $nf$ resonances with increasingly  higher $n$ occur as the ion charge increases. The above interpretation is supported by relativistic multiconfiguration Dirac-Fock (MCDF) calculations \cite{Parpia1996, Fritzsche2012a} as is demonstrated in Fig.~\ref{fig:strength} where we compare the calculated (squared modulus of the) electric-dipole matrix elements $\vert\langle 3d\vert\vert E1\vert\vert nf\rangle\vert^2$ with the strengths of the experimentally observed resonances from the fits in Fig.~\ref{fig:hires}. Since the (single-electron) wave functions of the $3d$ and $nf$ electrons are orthogonal to each other, we have chosen the (squared magnitude) of the electric-dipole matrix elements as this best reflects the probability for a $3d \to nf$ excitation and, hence, also the spatial overlap of these shells. As seen from Fig.~\ref{fig:strength}, the electric-dipole matrix elements become larger with increasing charge and decrease for larger $n$. We shall note, however, that the increase of the one-electron matrix elements reflects the overlap of the $3d$ and $nf$ orbitals and, hence, the contraction of the $nf$ wave functions but does not tell much about the resonance strength as observed experimentally. The latter one depends on the many-electron ionization and subsequent decay amplitudes and is much more difficult to compute reliably (cf.~Fig.~\ref{fig:xe5}).

To gain further insight into the formation of the resonances in Fig.~\ref{fig:hires}, one needs to analyse the
multiple Auger deexcitation processes which follow the  excitation or ionization of a $3d$ electron. Such a detailed analysis is beyond the scope of the present work. However, we can still compare the measured and calculated resonance structures by assuming that the branching ratio for triple ionization is the same for all resonances. This assumption appears to be well justified in view of a recent low-resolution study \cite{Schippers2014} where cross sections, also for double and quadruple ionization, were measured and where the resonance structures in the different final channels were found to be similar to one another. Moreover, we assume identical widths for all calculated resonances. Thus, all these resonances were represented by a Voigt line profile with a Lorentzian width of 0.6~eV and a Gaussian width of 0.16~eV.

Fig.~\ref{fig:xe5} compares the measured cross section with calculated data for the triple ionization of Xe$^{5+}$. To bring the theoretical data to the scale of experiment, we multiplied the theoretical (absorption) cross section by a (global) factor 0.3. This factor corresponds to a 30\% probability that Xe$^{8+}$ ions are produced by the Auger cascades that follow an initial $3d\to nf$ excitation in the primary Xe$^{5+}$ ion. In view of the crude assumptions made above, the agreement between the experimental and theoretical cross sections is quite satisfying. The largest deviation in the theoretical cross sections, when compared with experiment, occurs for the position of the $3d_{5/2}^{-1}\,5p\,4f$ resonances. It amounts to 2.8~eV and is in line with what we expect for a rather small-scale calculation as applied here to resonant photoionization of Xe$^{5+}$ with its $5s^2 5p$ ground configuration. In these computations, only the ionization via the $3d\to nf \;\, (4 \leq n \leq 8)$ excitations was considered as these channels are the most important ones for the observed resonance structures.

More extensive calculations employing large configuration expansions can be expected to substantially improve on the theoretical resonance energies, in particular, in the case of the $3d_j^{-1}\,4f$ resonances which are subject to comparatively strong correlation effects. Such large scale calculations would also reduce the currently too large (fine structure) energy splittings within these groups of resonances. The calculations would also have to account for photoionization of ions in long-lived metastable levels which are expected to have been present in the Xe$^{q+}$ ion beams \cite{Schippers2014}. More experimentally observed details, such as the irregularly strong $3d_{5/2}^{-1}\,5f$ resonance strength in the triple ionization cross section of Xe$^{4+}$ (Fig.~\ref{fig:hires}), can be addressed only by means of large-scale calculations which fully account for the Auger and radiative cascades following an initial $3d$ excitation. Until the present, however, such detailed computations of multiple cascades have been performed in some approximate manner only \cite{Kochur2009,Andersson2012}.

\section{Summary and Conclusions}

In summary, we have measured absolute cross sections for the triple ionization of multiply charged Xe$^{3+}$, Xe$^{4+}$, and Xe$^{5+}$ ions in the vicinity of the $3d$ ionization threshold. The measured cross sections exhibit a threshold due to the direct $3d$ ionization as well as strong resonances associated with $3d\to nf$ excitations and their subsequent autoionization. The number of these resonances increases with the charge state of the initial ion. This nicely demonstrates the formation of the $nf$ Rydberg resonances or, loosely speaking, shows how  the $nf$ wave functions \textit{collapse} into  the ionic core as $n$ increases. This conclusion is supported also by our atomic structure calculations. The present photoionization measurements, therefore, visualize directly the transition from a dilute spectator density of excited electrons into a regular shell structure. This may affect our understanding of chemical binding from a non-covalent association of electrons to covalent bonds as well as the behaviour of quantum dots and nanostructures. Quantum dots are often referred to as \lq\lq artificial atoms\rq\rq\ and their excitation (strength) strongly depends on the size and shape of the associated shell structure \cite{Kouwenhoven2001}, with applications in semiconductors, solar cells or quantum information \cite{Chang2007,Biolatti2002}.

\ack

This research was carried out at the light source PETRA\,III at DESY, a member of the Helmholtz Association (HGF). We would like to thank L. Glaser, F. Scholz, J. Seltmann, and J. Viefhaus for assistance in using beamline P04. This research has been funded in part by the German ministry for education and research (BMBF) under contracts 05K10RG1 and 05K10GUB within the \lq\lq Verbundforschung\rq\rq\ funding scheme.

\section*{References}


\begin{thebibliography}{10}

\bibitem{GoeppertMayer1941}
Goeppert-Mayer M 1941 {\em Phys. Rev.\/} {\bf 60} 184

\bibitem{Griffin1969}
Griffin D~C, Andrew K~L and Cowan R~D 1969 {\em Phys. Rev.\/} {\bf 177}

\bibitem{Lucatorto1981}
Lucatorto T~B, McIlrath T~J, Sugar J and Younger S~M 1981 {\em Phys. Rev.
  Lett.\/} {\bf 47} 1124

\bibitem{OSullivan1996}
O'Sullivan G, McGuinness C, Costello J~T, Kennedy E~T and Weinmann B 1996 {\em
  Phys. Rev. A\/} {\bf 53} 3211

\bibitem{Koizumi1997}
Koizumi T, Awaya Y, Fujino A, Itoh Y, Kitajima M, Kojima T~M, Oura M, Okuma R,
  Sano M, Seikioka T, Watanabe N and Koike F 1997 {\em Phys. Scr.\/} {\bf 1997}
  131

\bibitem{Bizau2001}
Bizau J~M, Cubaynes D, Esteva J~M, Wuilleumier F~J, Blancard C, Bruneau J,
  Champeaux J~P, Fontaine A~C~L, Couillaud C, Marmoret R, R{\'e}mond C, Hitz D,
  Delaunay M, Haque N, Deshmukh P~C, Zhou H~L and Manson S~T 2001 {\em Phys.
  Rev. Lett.\/} {\bf 87} 273002

\bibitem{Cummings2001}
Cummings A, McGuinness C, O'Sullivan G, Costello J~T, Mosnier J~P and Kennedy
  E~T 2001 {\em Phys. Rev. A\/} {\bf 63} 022702


\bibitem{Kjeldsen2002b}
Kjeldsen H, Andersen P, Folkmann F, Hansen J~E, Kitajima M and Andersen T 2002
  {\em J. Phys. B\/} {\bf 35} 2845

\bibitem{Habibi2009}
Habibi M, Esteves D~A, Phaneuf R~A, Kilcoyne A~L~D, Aguilar A and Cisneros C
  2009 {\em Phys. Rev. A\/} {\bf 80} 033407

\bibitem{Karaziya1981}
Karaziya R~I 1981 {\em Sov. Phys. Usp.\/} {\bf 24} 775

\bibitem{Connerade1982a}
Connerade J~P and Mansfield M~W~D 1982 {\em Phys. Rev. Lett.\/} {\bf 48}

\bibitem{Nuroh1982}
Nuroh K, Stott M~J and Zaremba E 1982 {\em Phys. Rev. Lett.\/} {\bf 49}
  862

\bibitem{Cheng1983}
Cheng K~T and {Froese-Fischer} C 1983 {\em Phys. Rev. A\/} {\bf 28}
  2811

\bibitem{Hansen1989}
Hansen J~E, Brilly J, Kennedy E~T and O'Sullivan G 1989 {\em Phys. Rev.
  Lett.\/} {\bf 63} 1934

\bibitem{Sonntag1984}
Sonntag B, Nagata T, Sato Y, Satow Y, Yagishita A and Yanagihara M 1984 {\em J.
  Phys. B\/} {\bf 17} L55

\bibitem{Arp1999}
Arp U, Iemura K, Kutluk G, Nagata T, Yagi S and Yagishita A 1999 {\em J. Phys.
  B\/} {\bf 32} 1295

\bibitem{Kivimaeki2000}
Kivim\"aki A, Hergenhahn U, Kempgens B, Hentges R, Piancastelli M~N, Maier K,
  R\"udel A, Tulkki J~J and Bradshaw A~M 2000 {\em Phys. Rev. A\/} {\bf 63}
  012716

\bibitem{Richter2007}
Richter T, Heinecke E, Zimmermann P, Godehusen K, Yal\ifmmode~\mbox{\c{c}}\else
  \c{c}\fi{}inkaya M, Cubaynes D and Meyer M 2007 {\em Phys. Rev. Lett.\/} {\bf
  98} 143002

\bibitem{Fritzsche2008}
Fritzsche S, J\"ank\"al\"a K, Huttula M, Urpelainen S and Aksela H 2008 {\em
  Phys. Rev.\/} {\bf 78} 032514

\bibitem{Maiste1980}
Ma\u{\i}ste A~A, Ruus R~{\'{E}}, Kuchas S~A, Karaziya R~I and {\'{E}}lango M~A
  1980 {\em Sov. Phys. JETP\/} {\bf 51} 474

\bibitem{Schippers2014}
{Schippers} S, {Ricz} S, {Buhr} T, {Borovik} Jr A, {Hellhund} J, {Holste} K,
  {Huber} K, {Sch{\"a}fer} H~J, {Schury} D, {Klumpp} S, {Mertens} K, {Martins}
  M, {Flesch} R, {Ulrich} G, {R{\"u}hl} E, {Jahnke} T, {Lower} J, {Metz} D,
  {Schmidt} L~P~H, {Sch{\"o}ffler} M, {Williams} J~B, {Glaser} L, {Scholz} F,
  {Seltmann} J, {Viefhaus} J, {Dorn} A, {Wolf} A, {Ullrich} J and {M{\"u}ller}
  A 2014 {\em J. Phys. B\/} {\bf 47} 115602

\bibitem{Viefhaus2013}
Viefhaus J, Scholz F, Deinert S, Glaser L, Ilchen M, Seltmann J, Walter P and
  Siewert F 2013 {\em Nucl. Instrum. Methods A\/} {\bf 710} 151

\bibitem{Mueller2009a}
M{\"u}ller A, Schippers S, Phaneuf R~A, Scully S~W~J, Aguilar A, Covington A~M,
  {\'A}lvarez I, Cisneros C, Emmons E~D, Gharaibeh M~F, Hinojosa G, Schlachter
  A~S and McLaughlin B~M 2009 {\em J. Phys. B\/} {\bf 42} 235602

\bibitem{Mueller2014}
M\"{u}ller A, Borovik Jr A, Buhr T, Hellhund J, Holste K, Kilcoyne A L D, Klumpp S, Martins M, Ricz S, Viefhaus J, and Schippers S 2014 {\em Phys. Rev. Lett.\/} in print

\bibitem{Larkins1977}
Larkins F~P 1977 {\em At. Data Nucl. Data Tables\/} {\bf 20} 311

\bibitem{Parpia1996}
Parpia F~A, Fischer C~F and Grant I~P 1996 {\em Comput. Phys. Commun.\/} {\bf
  94} 249

\bibitem{Fritzsche2012a}
Fritzsche S 2012 {\em Comput. Phys. Commun.\/} {\bf 183} 1525

\bibitem{Kochur2009}
Kochur A~G, Br\"{u}hl S, Petrov I~D and Mitkina Y~B 2009 {\em Eur. Phys. J.
  ST\/} {\bf 169} 51

\bibitem{Andersson2012}
Andersson E, Linusson P, Fritzsche S, Hedin L, Eland J~H~D, Karlsson L,
  Rubensson J~E and Feifel R 2012 {\em Phys. Rev. A\/} {\bf 85} 032502


\bibitem{Kouwenhoven2001}
Kouwenhoven L~P, Austing D~G and Tarucha S 2001 {\em Rep. Prog. Phys.\/} {\bf
  64} 701

\bibitem{Chang2007}
Chang C~H and Lee Y~L 2007  {\bf 91} 053503

\bibitem{Biolatti2002}
Biolatti E, {D'Amico} I, Zanardi P and Rossi F 2002 {\em Phys. Rev. B\/} {\bf
  65} 075306

\end{thebibliography}

\end{document}